%% file: DPF2019_Proceedy.tex
\documentclass[11pt]{article}
\usepackage[margin=1in]{geometry}
\usepackage{graphicx}

\input econfmacros.tex 

\def\Title#1{\begin{center} {\Large {\bf #1} } \end{center}}
\def\Author#1{\begin{center} {\normalsize {\sc #1} } \end{center}}
\def\Institution#1{\begin{center} {\normalsize {\it #1} } \end{center}}
\def\Abstract#1{\noindent {\normalsize {\bf Abstract:} {\normalfont #1}}}
\def\Conference{\vspace{4mm}\begin{raggedright} {\normalsize {\it Talk presented at the 2019 Meeting of the Division of Particles and Fields of the American Physical Society (DPF2019), July 29--August 2, 2019, Northeastern University, Boston, C1907293.} } \end{raggedright}\vspace{4mm}}

\begin{document}


%
%

\Title{ Quantum Tomography Measures Entanglement \\ in Collider Reactions}

\Author{John C. Martens, John P. Ralston, and Daniel Tapia Takaki}

\Institution{ Department of Physics \& Astronomy, University of Kansas, Lawrence KS, 66045}

\Abstract{ Entanglement in high energy and and nuclear reactions is receiving great attention. A proper description of these reactions uses density matrices, and the express of entanglement in terms of {\it separability}. Quantum tomography bypasses field-theoretic formalism to determine density matrices in terms of experimental observables. We review recent work applying quantum tomography to practical experimental data analysis. We discuss the relation between separability, as defined in quantum information science, and factorization, as defined in high energy physics. When factorization applies, it comes from using separable probes, which tomographically determine separable projections of entangled density matrices. }

\Conference

%
%

\section{Introduction}

{\it Entanglement} is receiving great attention in high energy and nuclear physics. There are good reasons: Entanglement is fundamental to this universe. Many papers follow textbooks defining entanglement for ``pure states,'' which are systems described by wave functions. Wave functions are actually an exceptional case, incapable of representing the general framework of quantum mechanics. Quantum states are generally described by density matrices. All inclusive reactions need a density matrix description. Yet physics history was first developed with wave function-based descriptions of non-relativistic exclusive reactions. That was transcribed to quantum field theory, conventionally organized to describe the most complicated systems conceivable, regardless of what will be observed. Using density matrices bypasses much of the unobservable formalism to describe exactly what is observable in the most simple way. 

Contrary to early lore, {\it everything quantum mechanics describes is observable} with density matrices. The notion that quantum systems are inherently unobservable was based on obsolete interpretations and formalism. {\it Quantum tomography} is a process by which high-dimensional information is built up by observing one-dimensional projections. The procedure is model-independent, and systematically transforms experimental data into the underlying components of an observed system's density matrix. Reference \cite{ourselves} discusses a practical, data-driven application which reconstructs the polarization density matrix of $Z$-bosons produced in high energy collisions. The entire superstructure of the traditional approach is bypassed. There are no structure functions, no field-theoretic $S$-matrix, no parton model, and no confusing dependence on coordinate frames. The numerical application leads to a convex optimization algorithm which has one global minimum, and no frustrating multiple minima. The method is so efficient that our supplemental Mathematica code goes straight from momentum 4-vectors to the hadronic density matrix in seconds. The density matrix then leads to true quantum-mechanical observables such as the {\it entanglement entropy}, which have been unavailable before quantum tomography became recognized. 

The thrust of reference \cite{ourselves} is practical data analysis, which transforms experimental data into the density matrix that produced it. Then the general features of entanglement {\it in terms of density matrices} casts a new light on hadronic physics and perturbative QCD. In particular, when a reaction conforms to the framework of {\it factorization}, it involves subsystems that are {\it separable} under the probe being used. Separability is one of the hallmark criteria of quantum information science. It is exceptional for interacting systems. Dynamically interesting systems are not separable, and that is the actual definition of {\it entanglement}. The focus of perturbative QCD on processes that factorize up to certain criteria is a focus on processes where entanglement certainly occurs, while the subprocess of interest is chosen to make it unobservable. We believe this is new.

\subsection{Background} 

A density matrix $\rho$ is an operator with positive eigenvalues, also called a {\it positive matrix.} Positive eigenvalues are real, Hermitian operators have real eigenvalues, hence $\rho = \rho^{\dagger}$. The expectation value of operator $A$ given density matrix $\rho$ is \begin{eqnarray}  <A> = {tr(\rho A) \over tr(\rho)} \rightarrow tr(\rho A). \label{one}\end{eqnarray} Here $tr$ stands for the trace. The first term on the right-hand side cancels the normalization of $\rho$, which depends on experimental selections. The last term applies with the convention $tr(\rho)=1$. Three textbook postulates for density matrices amount to one critical fact of positivity, which we will use below. 

The set of all operators of a given dimension and type form a vector space. The Hilbert-Schmidt inner product of operators $A$, $B$ is \begin{eqnarray} <A|B> =tr(A^{\dagger} B) \rightarrow tr(AB), \nonumber\end{eqnarray} the last assuming Hermiticity. Then $<A> =  tr(\rho A)$ acquires a geometrical meaning as the projection of vector $\rho$ onto vector $A$. Let $G_{\ell}$ be a set of orthonormal Hermitian operators, $<G_{\ell}|G_{k}> =tr(G_{\ell}G_{k})=\delta_{\ell k}$. The set forms a basis for the subspace it spans. Expanding $\rho$ in the basis gives \begin{eqnarray} & \rho = \sum_{\ell} \, |G_{\ell}><G_{\ell}|\rho>, \nonumber \\ & \rho = \sum_{\ell} \, <G_{\ell}>G_{\ell}. \label{rhonot} \end{eqnarray} Each independent observable $ <G_{\ell}>$ discovers one projection of the density matrix. It is appropriate to call each $G_{\ell}$ a {\it probe} of the system. Observing with a sufficient number of probes allows an arbitrarily complete reconstruction of the density matrix. This is {\it quantum tomography.}  

Eq. \ref{rhonot} might seem to require an infinite number of observables on the infinite dimensional spaces of quantum mechanics. Yet quantum mechanics is concerned with what is observable. The density matrix {\it of our description} of a system is exactly the density matrix that can be observed, just as Eq. \ref{rhonot} states. This is very powerful. There is a long tradition of classifying and modeling the properties of strongly interacting systems with many more theoretical inputs than experimental outputs. It is usually considered necessary, while strictly speaking, it is
completely unnecessary. The probes are chosen from known systems that have been understood, classified and re-arranged into orthogonal components. Then by what we call ``the mirror trick,'' the observed density matrix consists of just the same projections that are probed, just as Eq. \ref{rhonot} shows. A typical field theoretic description does just the opposite. It attempts to predict in advance infinitely many possible outcomes first, and project onto what is observed last. Quantum tomography is extremely efficient because it never needs to deal with what is not observed.

\subsection{The Born Rule}

Let $|e_{\alpha}>$ be the normalized eigenvectors of $\rho$, with eigenvalues $\rho_{\alpha}$. The ``spectral resolution'' of the operator is \begin{eqnarray} \rho= \sum_{\alpha} \, |e_{\alpha}>\rho_{\alpha}<e_{\alpha}|. \nonumber \end{eqnarray} Consider the expectation value of an operator $E_{\beta} =|e_{\beta}><e_{\beta}|$, for some value of $\beta$. It is \begin{eqnarray} <E_{\beta}> = tr(|e_{\beta}><e_{\beta}|\rho)=<e_{\beta}|\rho|e_{\beta}> = \rho_{\beta}, \nonumber \end{eqnarray} barring degeneracies. With $\rho$ being positive and $tr(\rho)=1$, the diagonal elements of $\rho$ in its eigenframe act like classical probabilities. Suppose $|f>$ is not an eigenvector of $\rho$, and $F=|f><f|$. Then \begin{eqnarray} <F> =tr(|f><f|\rho) = \sum_{\alpha} \, \rho_{\alpha}|<f|e_{\alpha}>|^{2}. \nonumber \end{eqnarray} This reproduces the Born rule probability to find $|f>$, given $|e_{\alpha}>$, summed over with classical probability weights to find each $|e_{\alpha}>$. 

In an exceptional case $\rho$ has $rank$-1, namely one non-zero eigenvector, denoted $|\psi>$, which defines the {\it wave function}. Eigenvectors have no definite phase, and a convention for normalization, explaining why wave functions have these features. Given $\rho \rightarrow |\psi><\psi|$, then $<A> \rightarrow <\psi|A|\psi>$. This, plus the Born rule specialized to rank-1, reproduces the rules of elementary quantum mechanics, exposed to refer to the rare case called the ``pure state''. There does not seem to be much experimental evidence for pure states, since interactions and entanglement do not preserve the concept. When $rank(\rho) \neq 1$ no wave function exists to replace the density matrix, which might be called the ``generic state.''

\subsection{Uses of a Density Matrix}

Once a density matrix has been tomographically reconstructed from experiment, it is the ideal summary of the system. Theory and experiment can meet in a common, well-defined framework that is as fundamental as quantum mechanics itself. 

Moreover, true quantum mechanical observables {\it not directly observed} can be computed. Here are three examples: \begin{itemize} \item The quantum probability of a normalized state $|f>$ is simply $<f|\rho|f>$. This concept is radically different from the classical probability of a state. In defining the classical probability, the state-space is discretized on some scale of resolution, and all possible distinct states are enumerated, as in classical statistical mechanics. As the resolution goes to zero there are infinitely many mutually-exclusive states. Quantum probability treats only orthogonal states as being mutually-exclusive. There are only $D$ orthogonal states on a $D$-dimensional space, so that quantum probability is inherently simpler, and also {\it not} correctly defined by beginning with classical probability. At the same time, quantum probability
rules allow composition of systems and subsystems, as in parton showers, which probe the evolution of density matrices, not classical distributions. \item The entropy $S$ of a density matrix $\rho$ is $$ S =-tr(\rho \log(\rho)).$$ When $\rho$ is equivalent to a pure state, then $S=0$, which is the minimum possible. When $\rho$ has no information, it equals the unit matrix times a normalization. Then $S =\log(D)$ for a $D \times D$ matrix. The entropy is a measure of the effective dimensionality of a system. When and if the entropy might show a significant change from a baseline, it is a signal that something significant is underway. \item The {\it projection postulate} appears in some presentations of beginning quantum mechanics. It states that upon making a measurement, a wave function will collapse to the state measured, and remain in that state until the next measurement. Since it refers to wave functions that are not general, and time evolution that is not general, it cannot be a consistent element of the theory. The literature on basic quantum mechanics\cite{Ballentine} recognizes this.  A better presentation recognizes that {\it in some cases} a projector $\pi_{1}$, $\pi_{1}^{2}=\pi_{1}$ might be measured. Suppose after that a different projector $\pi_{2}$ is measured, and nothing else evolves. If $[\pi_{1},\, \pi_{2}]\neq 0$, the result is different from measuring in the reverse order: \begin{eqnarray} &\rho \rightarrow \pi_{1}\rho \pi_{1} \rightarrow  \pi_{2}\pi_{1}\rho \pi_{1}\pi_{2} =\rho_{ 2\leftarrow 1}; \nonumber \\ & \rho \rightarrow \pi_{2}\rho \pi_{2} \rightarrow  \pi_{1}\pi_{2}\rho \pi_{2}\pi_{1} =\rho_{ 1\leftarrow 2} \neq \rho_{ 2\leftarrow 1}. \nonumber \end{eqnarray} This is called {\it outcome dependence}. As a rule no distribution of numbers (eigenvalues, etc.) associated with $\pi_{1}$ and $\pi_{2}$ can emulate outcome dependence. This underlies many so-called paradoxes of quantum measurement. The paradoxes come from invoking classical particles treated with classical probability that is not faithful to how probability works in quantum mechanics. A closely related error supposes that experiments measuring classical numbers could never capture the essence of quantum systems. But we have already explained how quantum tomography works. One simply needs {\it a sufficient number} of independent measurements to reconstruct a sufficiently informative density matrix, from which all of its possible observables can be computed. 

\end{itemize}

\subsection{Hadronic Reactions}

Let $p_{1}+p_{2} \rightarrow k_{1}+k_{2}+...+ X$ be a generic inclusive reaction. The conventional description in terms of $S$-matrix elements computes a cross section $d\sigma$: \begin{eqnarray} & d\sigma \sim \sum_{X} \, <p_{1}, \, p_{2}|_{in} \, k_{1}, \, k_{2}...X>_{out} <k_{1}, \, k_{2}...X|_{out} p_{1}, \, p_{2}>_{in} \, dLIPS, \nonumber \\ &  = \sum_{X} \, M_{p_{1}p_{2}k_{1}k_{2}...X}M^{\dagger}_{p_{1}p_{2}k_{1}k_{2}...X} dLIPS. \end{eqnarray} where $X$ is not detected, and $dLIPS$ absorbs the flux and phase space factors. The second line corrects common notational (and sometimes concept) errors in the first line, which come from assuming wave functions describe such systems. In fact, no wave function exists to describe unpolarized protons: ``Averaging over initial states'' and ``summing over final states'' are density matrix operations that actually contradict wave-function-based quantum mechanics. The second line implies no information, except to describe some multi-variate matrix of the form $MM^{\dagger}.$ Any matrix product $MM^{\dagger}$ has positive eigenvalues, and is a density matrix.

Since density matrices define quantum mechanics, there are no exception to the fact they always appear. Recall deeply inelastic scattering, the prototype inclusive reaction. Description assuming one-photon exchange processes was early separated into $d\sigma = L^{\mu \nu}W_{\mu \nu}$, suppressing phase space factors, where $L^{\mu \nu}$ and $W_{\mu \nu}$ are called leptonic and hadronic tensors, respectively. Both tensors are positive, so they are density matrices. The more general description is  $$ d\sigma = tr(\rho_{lep}\rho_{X}), $$ where $\rho_{lep}$ might be a more general lepton probe. From quantum mechanics $tr(\rho_{lep}\rho_{X}) =<\rho_{lep}>$. That is, {\it the scattered leptons} are observed, not the target. Decomposing $\rho_{lep}$ into orthogonal components, then (by the mirror trick) $\rho_{X}$ is a density matrix of projections onto the same subspace, which observes $\rho_{X}$. In terms of giving back from data what data measures, the parton model is an interpretation of a process that is conceptually {\it exact}, so long as it does not attempt to go beyond its own limits. 

The process of making structure functions for $\rho_{X}$ is superfluous for experimental purposes of {\it measuring} $\rho_{X}$ with the probe available. Structure functions will be part of a well-made theoretical model addressing what will be measured. Models do not need to predict everything. For example Bjorken scaling was predicted theoretically, at first on a tentative basis, and later by perturbative QCD, while the parton distributions (which are density matrices\cite{RS79}) are not predicted. The parton distributions represent a potentially infinite amount of information. To this day, theory predicts very little about strong interactions, compared to what {\it quantum tomography} has been doing, unrecognized, for more than 40 years. 


\section{Density Matrix Entanglement}

Schr\"odinger coined the word {\it entanglement} to expose the misunderstanding and misrepresentation of quantum probability. In beginning quantum mechanics two systems with variables $\vec x_{1}$, $\vec x_{2}$ are {\it not entangled} if the joint wave function $\psi(\vec x_{1}, \, \vec x_{2})  =\psi_{1}(\vec x_{1})\psi_{2}(\vec x_{2})$. Otherwise systems are entangled. To this day quantum probability is misrepresented in textbooks basing a presentation on classical probability defined by distributions.

In density matrix theory two systems $A, \, B$ are {\it separable} if the following criterion holds: \begin{eqnarray} \rho(A, \, B) = \sum_{\ell} \, P_{\ell}\rho_{\ell}(A) \otimes \rho_{\ell}(B). \label{separable} \end{eqnarray} Here $P_{\ell}>0$, and each factor $\rho(\ell)$ is positive. If systems are not separable they are {\it entangled}.  Let the Hamiltonian be $H(AB) =H_{A}+H_{B}+H_{AB}$, where $H_{AB}$ is the interaction term. Let $U_{A}(t) = e^{- \imath H_{A}t}$, and so on. Then \begin{eqnarray} U_{A}U_{B}\sum_{\ell} \, P_{\ell}\rho(1) \otimes \rho(2)U_{B}^{\dagger}U_{A}^{\dagger}= \sum_{\ell} \, P_{\ell}\rho_{\ell}(A) \otimes \rho_{\ell}(B). \nonumber \end{eqnarray} Thus when separability exists, it is invariant under time evolution neglecting interactions. Separable systems are typical candidates for zeroth-order perturbation theory. It should also be clear that maintaining separability in interacting systems is not a natural thing to expect. 

The criterion for density matrix entanglement did not appear until the paper of Werner\cite{Werner} in 1989, which came 63 years after quantum mechanics was discovered by Schr\"odinger in 1926. The fundamental role of the density matrix was suppressed for a long time. For example, in 1957 Fano\cite{fano} discovered quantum tomography and wrote Eq. \ref{rhonot}. It was largely ignored until the modern era of quantum computing resurrected it. Nowadays quantum tomography is an everyday term in quantum information science, while it is new and hardly explored in collider physics.

 \begin{figure}[ht]
\centering
\includegraphics[width=4.in]{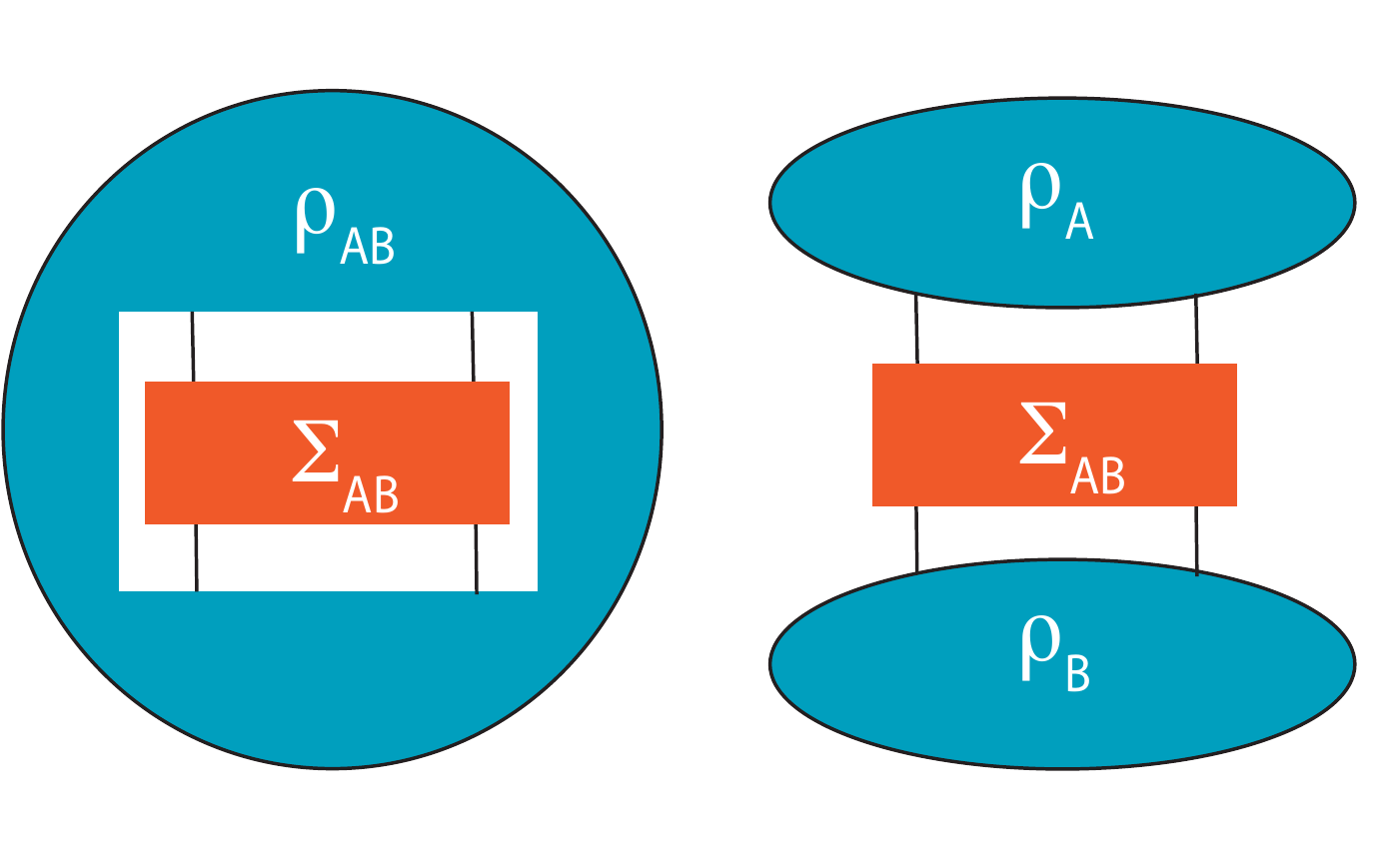}
\caption{Logic diagrams classifying systems. {\it Left:} An non-factorized, or entangled system, with a generic probe. Right: A factorized, or separable system {\it conditional upon} a special class of probe $\Sigma_{AB}$.  }
\label{fig:FactorSeparable}
\end{figure}

\subsection{It's All About the Probe} 
 
Figure \ref{fig:FactorSeparable} shows a diagram of probing separable versus entangled density matrices with probe scattering operators $\Sigma_{AB}$. The left hand panel indicates a generic, entangled system measured with a generic probe. There is no particular structure: ``everything interacts with everything else.''

Consider the outcome of certain special probes. Recall the elementary recipe that an operator $A$ acting on space $a$ is ``promoted'' to an operator acting on a product space $A \otimes B$ by replacing $A \rightarrow A \otimes 1_{B}$. If such an operator is measured in a joint density matrix $\rho(AB)$, then $tr(1_{B}\rho(AB))$ will convey no information about system $B$. In a scattering experiment we are interested in operators $\Sigma_{AB} \neq  A \otimes 1_{B} $,  $\Sigma_{AB} \neq 1_{A} \otimes B$, and indeed $\Sigma_{AB} \neq A \otimes B$, all of which would be incapable of measuring correlations. Compare $\Sigma_{AB} =\sum_{\ell} \,  A_{\ell} \otimes B_{\ell}. $ Let $A_{\ell}$ and $B_{\ell}$ be positive, and call $\Sigma_{AB}$ a {\it separable probe.} In particle physics positivity of the factors was assumed early on physical grounds seeking products of distributions and cross sections. That sort of operator can reveal correlations of a special kind, shown on the right-hand side of Figure \ref{fig:FactorSeparable}. The data from separable probes will be indistinguishable from measuring a density matrix that is {\it inherently} separable, Eq. \ref{separable}, for {\it all possible} measurements. We have the concept\footnote{To our knowledge, the quantum information community uses the word ``separable'' to refer to all possible measurements for the finite-dimensional systems of interest in the field. We do not find the concept of conditional separability there}.  of systems that are {\it separable conditional upon a particular probe}. 

\subsection{Factorization}

It is no great leap to go from Figure \ref{fig:FactorSeparable} to a ``factorization theorem'' represented by the same cartoons. Add external legs: The diagram {\it is} a basic factorization theorem, up to refinements for subtle (typically soft) perturbative interactions that do not actually factorize, but which can be computed and removed to make a theorem work out. The upshot is that {\it separability conditional upon a probe is factorization}, and vice-versa. Alternatively, {\it systems with probes that produce factorization are separable, and not entangled.} This is new.

The invariance of separability under subsystem time evolution allows perturbative QCD to speak of ``the'' parton distributions of separated systems, evolving order by order under their own internal Hamiltonian, {\it conditional upon a probe that makes the separation self-consistent.} We cannot expect every probe to produce a separable outcome, but when a probe is well-suited, one cannot detect the difference from intrinsically separable systems. Then for certain reactions in certain limits one can believe in ``parton distributions''. The parton distributions, and other density matrices, allow {\it some quantities} to be computed with the rules of classical probability, as if quantum mechanics never existed. In no event can {\it all quantities} defined in quantum mechanics be computed with classical probability. 


%
%
%

\section{Summary}

Inclusive reactions are described by density matrices. Quantum tomography circumvents unobservable superstructure to describe quantum mechanical systems in terms of how they are observed. Quantum tomography makes no models of unknown systems, while it is theoretically and computationally efficient. Our prototype example\cite{ourselves} includes computer code to go directly from lab-frame 4-vectors to density matrix elements. That can always be done. Once a density matrix has been tomographically reconstructed from data, every experimental outcome can be computed, to explore inherently quantum mechanical features coming from entanglement. 

Decades of beautiful calculations in perturbative QCD discovered a structure hidden in quantum mechanics. While entanglement of wave functions has always been recognized, the relation between factorization and separability of density matrices, along with its non-perturbative expression, seems to be new. These concepts have been a long time in formation. It might seem that the experts of quantum information science were far ahead of high energy physics. However the {\it perturbative} expression of factorization, with all the provisos of leading power, leading logs, etc. appeared in high energy physics many years before quantum information defined separability in 1989. Since separability defines entanglement, factorization makes a statement about entanglement, which has been done by avoiding entanglement.

\end{document}

%% file: econfmacros.tex
\def\beq{\begin{equation}}
\def\eeq#1{\label{#1}\end{equation}}
\def\eeqn{\end{equation}}

\def\beqa{\begin{eqnarray}}
\def\eeqa#1{\label{#1}\end{eqnarray}}
\def\eeqan{\end{eqnarray}}

\let\bar=\overbar

\def\Dslash{\not{\hbox{\kern-4pt $D$}}}
\def\dslash{\not{\hbox{\kern-2pt $\del$}}}

\def\msb{{\bar{\ssstyle M \kern -1pt S}}}

